\documentstyle[preprint,aps,prl]{revtex}

\begin{document}

\ifpreprintsty\else
\twocolumn[\hsize\textwidth%
\columnwidth\hsize\csname@twocolumnfalse\endcsname
\fi
\draft
\tightenlines
\preprint{ }
\title { Spin dynamics from time-dependent spin density-functional
theory}
\author { Zhixin Qian and Giovanni Vignale}
\address
{Department of Physics,  University of Missouri, Columbia, Missouri 65211}
\date{\today}
\maketitle
\begin{abstract}
We derive the spin-wave dynamics of a magnetic material  from
the time-dependent spin density functional theory in the linear response
regime.  The equation of motion for the magnetization includes, besides the static spin
stiffness,  a   ``Berry curvature"  correction and  a damping  term. A gradient
expansion scheme based on the homogeneous spin-polarized electron gas is proposed
for the latter two quantities,  and the first few coefficients of the expansion are
calculated  to  second order in the Coulomb interaction.  
\end{abstract}
\ifpreprintsty\else\vskip1pc]\fi
\narrowtext
\newcommand{\alda}{^{\scriptscriptstyle \rm ALDA}}
\newcommand{\qqp}{{\bf q}_{||}}
\newcommand{\ppp}{{\bf p}_{||}}
\newcommand{\kkp}{{\bf k}_{||}}
\newcommand{\qp}{q_{||}}
\newcommand{\pp}{p_{||}}
\newcommand{\kp}{k_{||}}

The study of the ground-state properties of magnetic materials within the
framework of spin-density functional theory (SDFT)  is by now a mature
field \cite{barth,rajagopal,sandratskii}.   By comparison, the  study of
excited-state properties is still in its infancy.
There has been great interest in recent years in deriving a closed equation of
motion for the magnetization starting from a first-principle description of
the electrons as itinerant particles
\cite{niu80,niu83,gebauer,antropov}, rather than from the
time-honored Heisenberg model of interacting local moments
\cite{herring,moriya,aharoni}. An alternative approach is to calculate  the
spectrum of spin excitations from  the imaginary part of the linear spin-spin
response function \cite{Savrasov}.    Our objective in this paper is to unify the two
approaches within the framework of  the {\it time-dependent} SDFT. We emphasize
new aspects of the physics beyond the adiabatic approximation (namely,  dissipation) 
as well as a practically workable computational scheme.

Elementary spin excitations in itinerant-electron magnets fall into two
groups  (i) Stoner excitations - in which a single electron quasiparticle is
spin-reversed  (ii) Spin waves.   Both types of excitations can be computed from the
linear spin-spin response function $\chi_{ij}({\bf r}, {\bf r}'; \omega)
=  i {\beta}^2 \int_0^\infty dt e^{i \omega t}
\langle [ \hat S_i ({\bf r} , t) ,
\hat S_j ( {\bf r}'
) ] \rangle$, ($\hat  S_i ({\bf r})$ is the the $i$ component of the
spin-density
operator, with $\beta = g e/(2 m c)$  and $\hbar =1$)  which determines
the  magnetization $m_{i}({\bf r},\omega ) = - \beta \langle
S_{i} ({\bf r} , \omega ) \rangle $
induced by an external magnetic field ${\bf B} ({\bf r}',\omega )$ at a
frequency $\omega$: \begin{eqnarray} \label{chi}
m_{i}({\bf r},\omega ) =\sum_j  \int \chi_{ij} ({\bf r} , {\bf r}' ; \omega )
B_j ({\bf r}' ; \omega ) d {\bf r}' .
\end{eqnarray}
Stoner excitations are distributed
along  branch cuts of this response function, while collective modes show
up as isolated poles in the complex frequency plane.

A time-dependent SDFT is ideally suited for calculating
$\chi_{ij}$ \cite{Gross}.   As a first step in this direction one solves  the
static Kohn-Sham equation \cite{kohnsham}, whose eigenfunctions and
eigenvalues   determine the exact equilibrium density and magnetization.
One then constructs the linear spin-spin response function
$\chi_{KS,ij}({\bf r} , {\bf r}' ; \omega ) $ of the Kohn-Sham (KS)
system.   Because this is a stationary noninteracting system, the
calculation can be carried out exactly \cite{grosskohn}.
Finally, the response of the physical
system is calculated as the response of the KS system  to an effective
time-dependent
field ${\bf B}_{eff}$ which includes, in addition to the external
field ${\bf B}$,  a many-body  ``exchange-correlation" field ${\bf B}_{xc}$.
When these ideas are cast in formulas one obtains the well-known
mathematical relation between the matrix inverses of the exact response
function and the KS response function, namely
\begin{eqnarray} \label{chi-1} [{\chi}^{-1}]_{ij} ({\bf r} ,
{\bf r}' ; \omega ) = [{\chi_{KS}}^{-1}]_{ij} ({\bf r} ,  {\bf r}'; \omega) -
f_{xc,ij} ({\bf
r} , {\bf r}' ; \omega ),
\end{eqnarray}
where $f_{xc,ij}({\bf r} , {\bf r}' ; \omega )$ is the tensor that connects
${\bf B}_{xc}$ to the  induced magnetization
\begin{eqnarray} \label{fxc} B_{xc,i}({\bf r},\omega ) =\sum_j  \int f_{xc,ij}
({\bf r} , {\bf r}' ; \omega )  m_j ({\bf r}' ; \omega ) d {\bf r}' .
\end{eqnarray}
Collective spin excitations  can then be obtained from the solution of the
eigenvalue problem
\begin{eqnarray}
\label{eigenvalueproblem}
\sum_j \int [{\chi^{-1}}]_{i j} ({\bf r}, {\bf r}' ; \omega )
m_j ({\bf r}', \omega ) d {\bf r}' = 0
\end{eqnarray}
where  $m_j ({\bf r}, \omega ) $ is the magnetization profile in the spin
wave.  The problem is to find the frequencies $\omega$ for which this
equation has nonvanishing solutions.   Because $\chi$ has both real and
imaginary parts these eigenfrequencies will be complex in general
and will determine both the dispersion ($Re~\omega$)  and the linewidth
($Im~\omega$) of  spin waves.

To appreciate the power of Eqs.~(\ref{chi-1}) and (\ref{eigenvalueproblem})
we now use them to ``derive" both the adiabatic spin dynamics
\cite{niu80,niu83,gebauer} and the Landau-Lifshitz (LL)
equation \cite{LL} in the linear response regime.  First of all, we  choose to
focus on the {\it transverse} part of
the response function, namely, the part that
describes the response to a magnetic field {\it perpendicular}  to the
ground-state magnetization.   It turns out that in a collinear magnet the
transverse response function is  rigorously decoupled from the longitudinal
one   in the absence of spin-orbit interactions.  In general,   this
decoupling is justified by the  difference between the time scales of
the longitudinal and transverse spin dynamics.   Our key assumption is that
both $\chi_{KS}^{-1}$ and $f_{xc}$ can be Taylor-expanded, at low
frequency, in powers of $\omega$:  this is justified because in a magnetic
material the scale of the frequency dependence of $\chi_{KS}^{-1}$ is set
by the
``Stoner-gap",   while
$f_{xc}$  is controlled by multiple electron-hole pair excitations, whose
spectral density is smooth. Keeping only the first
order term in the frequency  expansion of $\chi^{-1}$ we come to
 \begin{eqnarray}
\label{lowfrequencyexpansion}
[\chi ^{-1}]_{i j } ({\bf r}, {\bf r}' ; \omega ) = \alpha_{i j }  ({\bf
r}, {\bf r}' )
 + i \omega  \tilde \Omega_{i j} ({\bf r}, {\bf r}' )
\end{eqnarray}
where $ \alpha_{i j} ({\bf r }, {\bf
r}') \equiv
[\chi ^{-1}]_{i j } ({\bf r}, {\bf r}' ; \omega = 0 )$ is the (symmetric)
{\it spin
stiffness} tensor, given by the second derivative of the  ground-state energy with
respect to the magnetization,
and
\begin{eqnarray}
\label{derivative}
\tilde \Omega_{i j } ({\bf r}, {\bf r}' ) \equiv \lim_{\omega \rightarrow 0}
\frac{\partial Im [\chi ^{-1}]_{ij} ({\bf r} , {\bf r}' ;
\omega )}{\partial \omega} ,\end{eqnarray}
with the derivative taken along the real frequency axis.
($\tilde \Omega_{ij}$ is purely real because the derivative of $Re \chi$
vanishes at $\omega = 0$.)
Notice that, by definition,
\begin{eqnarray}
\int \alpha_{i j} ({\bf r} , {\bf r}' ) m_j ({\bf r}' ) d { \bf r}'
=  \frac{\delta E [ {\bf m }] } {\delta m_i ({\bf r} )},
\end{eqnarray}
where $E [ {\bf m }]$ is the ground-state energy regarded as a functional of
${\bf m} ({\bf r})$ \cite{footnote1}.  The tensor $\tilde \Omega_{i j }$
can be split into antisymmetric and symmetric components as follows
\begin{equation} \label{tildeomega}
\tilde \Omega_{ij} ({\bf r} , {\bf r}' ) = \Omega_{ij} ({\bf r} , {\bf r}' ) +
\gamma_{ij} ({\bf r} , {\bf r}' ),
\end{equation}
where $\Omega_{ij} ({\bf r} , {\bf r}' )  = -\Omega_{ji} ({\bf r}' , {\bf
r} ) $ and
$\gamma_{ij} ({\bf r} , {\bf r}' )  =  \gamma_{ji} ({\bf r}' , {\bf r} ) $.
Therefore, substituting
(\ref{lowfrequencyexpansion}) into (\ref{eigenvalueproblem}), and switching to
a real time representation with the substitution $-i \omega \to \partial
/\partial t$,
we obtain the equation of motion
 \begin{eqnarray}
\label{equationofmotion} \sum_j \int d {\bf r}' [ \Omega_{ ij} ({\bf r},
{\bf r}')
 + \gamma_{ij} ({\bf r}, {\bf r}' ) ]
\frac{\partial m_j ({\bf r}' ,t)}{\partial t}
= \frac{\delta E [{\bf m }]}
{\delta { m_i} ({\bf r}, t)}.
\end{eqnarray}

 This equation reduces to the Niu-Kleinman  adiabatic equation of motion
\cite{niu80} if the
symmetric tensor $\gamma_{ij}$  is neglected and the antisymmetric tensor
$\Omega_{ij}$ is identified with the ``Berry curvature" .  Indeed, after
a lengthy but straightforward calculation  we can show that
 \begin{equation} \label{Berrycurvature} \Omega_{ij} ({\bf
r} , {\bf r}') = -2   Im \left \langle {\partial \psi [{\bf m}] \over
\partial m_i
({\bf r}) }  \biggl \vert {\partial \psi [{\bf m}] \over \partial m_j ({\bf
r}') }
\right \rangle \end{equation}
where $ \psi [{\bf m}] $ is the ground-state wave function regarded as a
functional of ${\bf m}$ \cite{footnote1}.  Thus, the antisymmetric part of
Eq. (\ref{derivative})
is equivalent to  Eq. (\ref{Berrycurvature}): the former is, however,
 more amenable to approximation and computation.

The symmetric part of Eq. (\ref{derivative}),  $\gamma_{ij}$,  is responsible for {\it
dissipation} as one can immediately verify by calculating the rate of entropy
production at  temperature $T$:
 \begin{eqnarray} \label{entropyproduction} T {d S \over d t} &=&
-\int {\delta  E [{\bf m }] \over \delta {\bf m} ({\bf r}, t) } \cdot
{\partial {\bf m}
({\bf r}, t) \over \partial t} d {\bf r}   \nonumber  \\
&=& - \sum_{ij}\int d {\bf r} \int d {\bf r}'
{\partial {\bf m}_i ({\bf r}, t) \over \partial t} \gamma_{ij} ({\bf r},
{\bf r}' )
{\partial {\bf m}_j ({\bf r}', t) \over \partial t}.  \nonumber  \\
\end{eqnarray}
Not surprisingly, this term is absent in a purely adiabatic theory
such as that of Ref. \cite{niu80}.

 We now turn to the task of approximating the right hand side of Eq.
(\ref{derivative}).  A classic approximation scheme is provided by the
 gradient expansion \cite{Perdew}.  In this scheme one assumes that  the
two-point function $\tilde \Omega_{ij}({\bf r},{\bf r}')$ is a  short ranged function of
the distance $|{\bf r}- {\bf r}' |$.  It is then permissible, if the density and magnetization
are  slowly varying, to  expand  $\tilde \Omega$  as
\begin{eqnarray}
\label{gradientexpansion}\tilde  \Omega_{ij } ({\bf r}, {\bf r}' ) &=&
\tilde \Omega_{0,ij }[n({\bf r}),m({\bf r})]  \delta ({\bf r}- {\bf r}' )
\nonumber \\ &+& \tilde \Omega_{2,ij } [n({\bf r}),m({\bf r})]  {\bf
\nabla_r}\delta ({\bf r}- {\bf r}' ) \cdot {\bf \nabla_{r'}} + ~...
\end{eqnarray}
where $\tilde \Omega_{0,ij} [n,m]$ and $\tilde \Omega_{2,ij}[n,m]$ are
the coefficients of $q^0$ and $q^2$ respectively  in the small-$q$ expansion
of $\tilde \Omega^{hom}_{ij}({\bf q}) \equiv \int \tilde
\Omega^{hom}_{ij}({\bf r} - {\bf r}')e^{-i {\bf q} \cdot ({\bf r} - {\bf r}')} d
{\bf r}$ in a homogeneous electron gas of density $n$ and magnetization
${\bf m}$.

We are now in a position to prove that the standard LL equation \cite{LL}
is simply  the zero-order approximation   (i.e., the local density
approximation)  in the gradient expansion for the Berry curvature.
To this end, we consider a homogeneous spin-polarized  electron gas with
the same ground state density $n$ and magnetization ${\bf m}_0 $ as those of
the real system at point ${\bf r} $.    The homogeneous magnetization  is
maintained by an external fictitious  magnetic field ${\bf B}_0$
 parallel
to  ${\bf
m}_0 $,
\begin{eqnarray}
B_{0} = \left ( \frac{\partial \epsilon (n, m)}{\partial m } \right
)_{m={m_0}}
\end{eqnarray}
where $\epsilon (n,m)$  is the energy density of the homogeneous electron
gas of density $n$ and magnetization $m$ \cite{footnote2}.
The transverse spin-spin response function of this system  at $q=0$  is
\begin{equation} \label{chiq0}
[\chi^{-1}]_{ij}(q=0,\omega) = { \omega_0 \delta_{ij} + i \omega
\epsilon_{ij} \over
\beta m_0}
\end{equation}
where $\omega_0 = \beta B_0$ and $\epsilon_{ij}$ is the
two-dimensional Levi-Civita  tensor with the
cartesian indices $i,j$ being orthogonal to the direction of  ${\bf m}_0$.
Thus, for the homogeneous electron gas, $\tilde \Omega_{ij}(q=0) =
\epsilon_{ij}/(\beta m_0)$ (see Eq.~ (\ref {derivative})) , and then, from
Eqs. (\ref{tildeomega}) and (\ref{gradientexpansion}), we see that  the
local density approximation takes the form
 \begin{eqnarray}
\label{gradientexpansion2} \Omega_{ij } ({\bf r}, {\bf r}' )  \simeq
\epsilon_{i j} \frac{1}{\beta m_0 ( {\bf r } )}  \delta ({\bf
r}- {\bf
r}') .
\end{eqnarray}
With this approximation Eq.~(\ref{equationofmotion}) reduces to the linearized
LL equation,
\begin{eqnarray}
\label{ll}
\frac{\partial m_i ({\bf r} , t )}{\partial t} = - \sum_j \beta m_0
\epsilon_{i j}
\frac{\partial E [{\bf m} ] }{\partial m_j ({\bf r}) } .
\end{eqnarray}
Notice that the dissipative  $\gamma_{ij}$ is exactly zero at
this  order of approximation. Thus, the gradient expansion for $\gamma_{ij}$
begins with a  second order term
 \begin{eqnarray}
\label{gradientexpansion3} \gamma_{ij } ({\bf r}, {\bf r}' ) =
 \gamma_{2,ij }
({\bf r})  {\bf
\nabla_r}\delta ({\bf r}- {\bf r}' )  \cdot {\bf \nabla_{r'}} + ~...  .
\end{eqnarray}
This makes physical sense because a global rotation of the spins must be
rigorously undamped  in the absence of spin-orbit interactions.

The remainder of this paper is  devoted to the
calculation  of the leading gradient corrections to the LL
equation.     After lengthy calculations, which will be
described in detail elsewhere, we obtain the small-$q$ expansion of the
transverse spin-spin response function of the  homogeneous
spin-polarized electron gas:
\begin{eqnarray} \label{chi-1qomega}
&&[\chi^{-1}]_{ij} ({\bf q}, \omega)  =  [\chi^{-1}]_{ij} ( 0 , \omega ) +
\frac{n q^2}{4 m m_0^2}  \delta_{ij}    \nonumber  \\
 && +  \frac{q^2}{(2 m m_0)^2}   \biggl [
\frac{4 m \langle \hat T_{\downarrow} - \hat T_{\uparrow} \rangle }{3 V}
\frac{\omega _0 \delta_{ij} - i \omega  \epsilon_{ij}}
{ \omega^2 -{\omega_0}^2 }   \nonumber
\\
&& + \frac{ 2 \pi  n^2( 2 g (0) -1 ) }{3 m a_0}   \left(
\frac{\delta_{ij} - i \epsilon_{ij} }{( \omega - \omega_0 )^2 }
+ \frac{\delta_{ij}+ i \epsilon_{ij}}{(\omega + \omega_0 )^2 } \right)
\nonumber \\
&& +  \frac{F_{-+} (\omega ) ( \delta_{ij} - i \epsilon_{ij})}
{ (\omega - \omega_0 )^2 }
+ \frac{F_{+-} (\omega ) ( \delta_{ij} + i \epsilon_{ij})}
{( \omega + \omega_0 )^2 }  \biggr ]      \nonumber  \\
\end{eqnarray}
where $V$ is the volume and $a_0$ is the Bohr radius,
 $\hat T_\uparrow$ and $\hat T_\downarrow$ are the kinetic energy
operators associated with up-spin and down-spin electrons respectively,  the
angular brackets denote the ground-state or thermal ensemble average,
$g (0)$ is the pair correlation function at zero
separation,
and  $F_{+-} ( \omega )$ is a four-point response functions,  defined as
\begin{eqnarray}
\label{fpm}
F_{+-} (\omega ) &=& F^*_{-+} (- \omega ) = \frac{1}{3V^3 }  \sum_{{\bf k},
{\bf k}'} v({\bf k} ) v( {\bf k}' ) {\bf k} \cdot {\bf k}'       \nonumber
\\
& & \langle \langle \hat S_+ (- {\bf k} ) \hat  \rho ({\bf k} );
\hat S_-({\bf k}' ) \hat \rho (-{\bf k}' ) \rangle \rangle _\omega.
\end{eqnarray}
$\hat S_\pm ({\bf k} ) = \hat S_x({\bf k} )  \pm i \hat S_y ({\bf k} ) $
are spin-density fluctuation operators, $v({\bf k}) = 4 \pi e^2/k^2$, and $\hat  \rho
({\bf k} )$ is the density
fluctuation operator. The ``Zubarev product" is defined as $\langle \langle
\hat A;
\hat B \rangle \rangle_\omega \equiv -i \int_0^\infty d t e^{i \omega t}
\langle
[\hat A(t), \hat B] \rangle$.
Taking the small $\omega$ limit of this expression we obtain the
coefficients of
the gradient expansion for $\Omega$ and $\gamma$ as follows:
\begin{eqnarray} \label{omega2}
\Omega_{2,ij}   &=&   \epsilon_{ij}   \frac{  1 } { (m m_0 \omega _0 )^2 }
\biggl [ \frac{m \langle T_{\uparrow} - T_{\downarrow} \rangle }
{3 V }  + \frac{ 2 \pi  n^2
( 2 g ( 0 ) - 1 )}
{3 m a_0 \omega_0 }                                 \nonumber \\
& & \hspace{0.8 cm} + \frac{Re F_{+-} (0)}{ \omega_0}
- \frac{ 1}{ 2  }  \frac{\partial  Re F_{+-} (\omega )}
{\partial \omega}\biggr |_{\omega =0} \biggr ]
\end{eqnarray}
and
\begin{eqnarray} \label{gamma2}
\gamma_{2,ij}= - \frac{  \delta_{ij}} { 2 (m  m_0 \omega_0 )^2 }
\lim_{\omega \rightarrow 0 } \frac{ Im F_{+-} (\omega ) }
{\omega }
\end{eqnarray}
 Since the long wavelength spin wave frequency in a ferromagnet is proportional
to $q^2$, the above results indicate that the gradient corrections to
$\Omega$ and
$\gamma$ will affect the dispersion and damping of ferromagnetic spin waves
beginning at  order $q^4$.

Equations ~(\ref{omega2}) and (\ref{gamma2}) contain both ground-state
(thermal ensemble) averages, such as $\langle T_{\uparrow} - T_{\downarrow}
\rangle$ and $g ( 0 )$ and the dynamical response
function   $F_{+-}(\omega)$ which depends on the spectrum of excited
states.  The
former can be calculated to a high degree of accuracy by
variational
and diffusion Monte Carlo techniques \cite{Ceperley};  the major challenge
lies therefore in the
calculation of $F_{+-} (\omega )$.  The form of Eq.~(\ref{fpm}) suggests
that we
evaluate $F_{+-}(\omega)$  to {\it second order} in the Coulomb interaction:
this is accomplished by substituting the noninteracting expression for the four
point  response funtion $\langle \langle \hat S_+ (- {\bf k} ) \hat  \rho
({\bf k} );
\hat S_-({\bf k}' ) \hat \rho (-{\bf k}' ) \rangle \rangle _\omega$.  Even such an
approximate calculation turns out to be very difficult, but we have been
able to establish analytically the  limiting forms of the  imaginary part
of $F_{+-} (\omega )$ for high  and low frequency.
At low frequency ( $\omega << E_{F \uparrow} - E_{F
\downarrow}$, where $E_{F
\uparrow }$ is the largest of the two Fermi energies) we find
\begin{eqnarray}
\label{ImFlow} Im {F_{+-}}  (\omega ) = - \frac{ m^2  \Gamma
\omega (\omega^2 + 4 \pi^2 k_B^2T^2)}{36 \pi^3 k_{F \uparrow} {a_0}^2},
\end{eqnarray}
where $k_B$ is the Boltzmann constant and  $T$ is the temperature.
The dimensionless coefficient $\Gamma$ is the sum of
``direct" and ``exchange" terms:  $\Gamma = \Gamma^{(D)} + \Gamma^{(E)}$ which
we report separately for reasons that will become clear in the following:
\begin{eqnarray} \label{gammas}
\Gamma^{(D)} &=& \frac{2 \lambda}{1 -\lambda^2} +\frac{\theta (3 \lambda -
1 ) (3 \lambda - 1 )}{2 \lambda (1 - \lambda )},
\nonumber \\
\Gamma^{(E)} &=& \frac{1}{ 2 }  \ln \frac{1 + \lambda}{1 - \lambda}
- \frac{\theta  (3 \lambda -1)}{2 \lambda} \ln \frac{2 \lambda}
{1 - \lambda} ,
\end{eqnarray}
where $\lambda \equiv (1 - \zeta)^{1/3}/(1 + \zeta)^{1/3}$,
and  $\zeta \equiv (n_{\uparrow} - n_{\downarrow} )/n $ is the degree of spin
polarization.
Eq.~(\ref{ImFlow}) yields  the damping tensor according to Eq.
(\ref{gamma2}):
\begin{eqnarray}
\gamma_{2,i j}   = \delta_{i j} \frac{ \Gamma }{2 \pi k_{F \uparrow}(3 m_0
\omega_0
a_0 )^2} ( k_B T)^2.    \end{eqnarray}
Note that the dissipation vanishes as $T^2$ for
$T \to 0$ \cite{Halperin}.

To calculate the correction to the Berry curvature we also need the real
part of
$F_{+-}(\omega)$.  To this end, we make use of the Kramers-Kr\"onig dispersion
relation
\begin{eqnarray} \label{KK}
Re F_{+ -} ( \omega ) = \frac{1}{\pi}
\int_{- \infty}^{\infty} d \omega ' \frac{Im F_{+ -}
( \omega ' ) } {\omega ' - \omega }
\end{eqnarray}
and use for $Im F_{+ -}(\omega)$ at finite frequency the mode-decoupling
approximation of Ref. \cite{nifosi} :
 \begin{eqnarray} \label{modedecoupling}
&& {1 \over V^2} Im \langle \langle S_{+} (- {\bf k} ) \rho ({\bf k}) ;
S_{-} ({\bf k}' ) \rho (- {\bf k}' ) \rangle \rangle _\omega \simeq
\nonumber  \\
&&  - g_x \frac{\delta_{{\bf
k} , {\bf k}' } } {\pi} \int_0^{\omega}  Im  \chi_{nn}
 ({\bf k}, \omega - \omega ' )     Im  \chi_{+ -} (-{\bf k} , \omega ' ) d
\omega '
\nonumber \\
 \end{eqnarray}
where $\chi_{nn}({\bf k},\omega) = V^{-1} \langle \langle \rho({\bf
k});\rho(-{\bf k}) \rangle \rangle_\omega $ is the density-density response
function, and the factor $g_x = (\Gamma^{(D)}+\Gamma^{(E)})/ \Gamma^{(D)}
$ is used to include the exchange contribution and assure the correct behavior in
the most important  low- frequency limit.

Carrying out the calculations, we find that the last two terms on the right
hand
side of
Eq.~(\ref{omega2}) are 
\begin{eqnarray}
Re F_{+-} (0) = - g_{x} \frac{{k_{F \uparrow}}^3 E_{F \uparrow}}{3 \pi^4
{a_0}^2} P (\zeta) ,
\end{eqnarray}
and
\begin{eqnarray}
\frac{\partial Re F_{+-} (\omega )}{\partial \omega } \biggr |_{\omega =
0}  = - g_{x} \frac{{k_{F \uparrow}}^3 }{ 3 \pi^4 {a_0}^2 } Q (\zeta ) ,
\end{eqnarray}
where $ P (\zeta ) $ and $Q (\zeta ) $ can be very accurately
parametrized as
\begin{eqnarray}
P (\zeta ) = 1.9606 - 3.5 \zeta - 1.4 \zeta ^2 \ln \zeta
+ 2.08 \zeta ^2 ,
\end{eqnarray}
and
\begin{eqnarray}
Q (\zeta ) &&~=1.18 \zeta - 0.186 \zeta^2 - 0.842 \zeta^3 \nonumber \\
&&~- (0.045 \zeta  - 1.49 \zeta^2)  \ln \zeta.
\end{eqnarray}

Finally, we provide practical approximate expressions for the ground-state
averages appearing in Eq.~(\ref{omega2}).    In the absence of detailed Monte
Carlo results for intermediate polarization we suggest simple interpolation
formulas between the paramagnetic ($\zeta = 0$) and the fully spin-polarized
($\zeta = 1)$ state.  For the zero-separation pair-correlation function
$g(0)$, a reasonable interpolation  is $g(0) = g_0 (0) (1 - \zeta^2 )$ where
$g_0 (0)$ is the paramagnetic state value, which is given in Ref.
\cite{ortiz}. As for $\langle T_{\uparrow} - T_{\downarrow} \rangle$,
we  propose a linear interpolation for the correlation contribution:
\begin{eqnarray}
\langle T_{\uparrow} - T_{\downarrow} \rangle =
\langle T_{\uparrow} - T_{\downarrow} \rangle _0
- N \frac{d ( r_s \epsilon_c ) }{d r_s } (\zeta = 1 )  \zeta,
\end{eqnarray}
where  $\epsilon_{c} $ is the correlation energy  per particle, $r_{s}$ the
Wigner radius, and $\langle T_{\uparrow} - T_{\downarrow} \rangle _0
= ({k_{F \uparrow}}^5
- {k_{F \downarrow}}^5 )/(20 \pi^2 m )$,
the difference of kinetic energies of the noninteracting systems.

The generalized Landau-Lifshitz equation (\ref{equationofmotion})
is the central result of this paper. It  includes both   the
adiabatic spin dynamics and the conventional Landau-Lifshitz equation
as special cases.
It reduces to the adiabatic spin dynamics if the damping tensor $\gamma $ is
neglected.  It further reduces
to conventional Landau-Lifshitz equation if the gradient corrections to
$\Omega$ are  neglected.
In addition, we have developed a density-functional scheme for the systematic
calculation of  $\Omega$ and $\gamma$.  The analytical expressions for the
damping tensor and  the Berry curvature make the
application of the new equation of motion to spin dynamics of ferro- and
antiferromagnetic systems quite
promising.   If greater accuracy is required, one can  revert to the
full-fledged linear response formalism, in which $\chi^{-1}_{KS}$ is treated
exactly and only $f_{xc}$ is approximated.

We gratefully acknowledge support from   NSF grant No. DMR-0074959 and
Research Board grant No. URB-00-029 at the University of Missouri.


\begin{references}
\bibitem{footnote1}  The precise definition is $E[{\bf m}] = min_{\psi \to
{\bf m} ({\bf r})} \langle \psi| \hat H |\psi \rangle$ where $\hat H$ is
the hamiltonian and the minimum is sought among  the  wave
functions that yield the magnetization  ${\bf m} ({\bf r})$.  This
constrained search also yields the ground-state wave function as a
functional of   ${\bf m} ({\bf r})$.
\bibitem{footnote2} Thus, we do not consider the possibility of
spontaneous ferromagnetism in the homogeneous electron gas. This is
predicted to occur at extremely low densities  \cite{ortiz},
and is of no interest here.
\end{references}
\end{document}